\documentclass[letterpaper]{article}

\usepackage[sc]{mathpazo} 
\usepackage[T1]{fontenc} 
\usepackage{graphicx}

\usepackage[hmarginratio=1:1,top=32mm,columnsep=20pt]{geometry} 
\usepackage[bookmarks=true,bookmarksnumbered=true,bookmarksdepth=1,pdfstartview=FitH]{hyperref} 

\usepackage[hang, small,labelfont=bf,up,textfont=it,up]{caption} 
\usepackage{booktabs} 
\usepackage{multicol}

\usepackage{abstract} 
\renewcommand{\abstractnamefont}{\normalfont\bfseries\normalsize} 
\renewcommand{\abstracttextfont}{\normalfont\small} 

\usepackage{sectsty} 
\sectionfont{\large}
\subsectionfont{\normalsize}
\subsubsectionfont{\rm}

\usepackage[bottom]{footmisc}

\newenvironment{biblist}{%
  \begin{list}{}{%
    \setlength{\labelwidth}{0pt}%
    \setlength{\labelsep}{1em}%
    \setlength{\leftmargin}{1em}%
    \setlength{\itemindent}{-1em}%
  }
}{\end{list}}

\newcommand{\al}{\alpha}
\newcommand{\iid}{\stackrel{{\rm iid}}{\sim}}
\newcommand{\vol}{{\rm vol}}

\title{\vspace{-15mm}On Sampling Social Networking Services} 

\author{
Baiyang Wang\thanks{Bachelor of Science (Actuarial Science) Candidate in the Department of Statistics and Actuarial Science, the University of Hong Kong. Email: \href{mailto:pakyeung@hku.hk}{pakyeung@hku.hk}. The author thanks all the help from Professor Mark S. Handcock in UCLA.}
}

\begin{document}

\maketitle 
\vspace{-2em}

\begin{abstract}
This article aims at summarizing existing methods for sampling social networking services and proposing a faster confidence interval for related sampling methods. Social networking services (SNSs), such as Facebook and Twitter, are an important part of the current Internet culture. Collecting samples from these networks, therefore, is necessary for learning more about sociological or cultural issues. However, typical sampling methods for networks, such as node-based or link-based methods, are not always feasible for social networking services. Alternate approaches such as snowball sampling or random walk (RW) are applied to gather information from social networking services more efficiently. Thus it is beneficial to compare various sampling approaches for SNSs under different circumstances.\\

Making statistical inference from the gathered information constitutes another problem, including the determination of the sampling probabilities and the estimation of related attributes for sampled networks. Many valuable approaches have been proposed and refined, yet their numerical properties have not been studied sufficiently, some of which will be shown in the following. Simulations will be carried out in terms of respondent driven sampling (RDS), which has many variations and comprises several important network sampling methods. Although it was initially invented to detect hidden population in society, it has been generalized and widely discussed in the related literature. Based on these existing methods, the author also proposes some modification of existing estimation methods and the construction of a faster confidence interval which has not been covered in the current literature.
\end{abstract}

\setlength{\parskip}{1em}

\section{Introduction}
Networks are an important natural phenomenon, and much efforts have been devoted to study their properties. In this article, we mainly consider social networks joined by different people, and we will pay particular attention to the social networks formed in online social networking services. For general social networks, the reader can refer to Wasserman and Faust (1994), which provides a comprehensive list of their properties. To obtain knowledge of these properties requires sampling from the networks, which have long aroused the interest of researchers, and some early work can be traced back to snowball sampling (Goodman, 1961). Generally, the sampling schemes for social networks can be very different from independent sampling, as it is very difficult to obtain independent samples from the networks and it is often necessary to reach one person from another. Indeed, many approaches of this kind ({\it chain-referral} or {\it crawling} methods) have been proposed for sampling social networks, such as BFS sampling, DFS sampling and random walk, just to name a few. Section 2 will be devoted to providing a list of common sampling methods for SNSs and noting the possible issues for consideration for each method.

As is mentioned, the network samples may not be independently distributed and to adjust the bias, one should apply weighted estimators such as the Hansen-Hurwitz estimator in Section 4.1 (Hansen and Hurwitz, 1943). Even then, for sampling methods without replacement, the sampling probabilities need to be estimated to obtain the numerical values of weighted estimators, which will be discussed in Section 3. Given these estimations, comparisons between different network sampling methods, especially variations of RDS methods, are feasible, which will be the main topic of Section 4, including comparisons between one-node-at-a-time (RW type) and multiple-nodes-at-a-time (RDS type) sampling, with-replacement and without-replacement ({\it traversal}) sampling and coverage probabilities for different confidence intervals for the attributes of sampled social networks, together with a new confidence interval which has similar performance with previous ones and lower computational costs. The effect of population size on network sampling and the homogeneity/heterogeneity between networks with the same properties will also be briefly discussed. The simulated networks will be based on the commonly used exponential random graph models (ERGMs) (Snijders, 2002). Section 5 discusses the works in the related literature and concludes the article.


\section{Current Network Sampling Methods}

\subsection{Node-based Network Sampling}
In certain situations, sampling independently from the nodes in a network is possible, and then we can apply simple node-based sampling, i.e. to randomly collect information from a group of people on SNSs and keep the links between them. Although this method generally preserves the topological structures of a social network (Lee et al., 2006), it is not easy to implement under many circumstances. While some available SNS datasets such as complete data sets for Facebook users in Harvard (Lewis et al., 2008) and Caltech (Traud et al., 2008) make this kind of sampling possible, online social networks are highly dynamic, and typical SNSs such as Facebook can only generate a small sample each time for a given group of people, which make the node-based sampling method difficult for SNSs. Moreover, such complete data sets also trigger privacy concerns.

Apart from that, there is an easier way to acquire a uniform random sample across giant online social networks, which is to generate random user IDs uniformly and then reject the IDs that do not match any user. It can be easily proved that this kind of rejection sampling generates the same results as uniform random sampling without replacements (Gjoka et al., 2010). This approach was previously available for Facebook; yet now, Facebook has stopped using number IDs, and this approach appears be less practical.

\subsection{Link-based Network Sampling}
Apart from node-based sampling, it is also possible to randomly sample from the links inside the networks and keep the nodes attached to them. The difference between node-based sampling and link-based sampling is illustrated as follows:
\newpage
\begin{figure}[h]
\begin{center}
    \includegraphics[width=2.5in,height=1in]{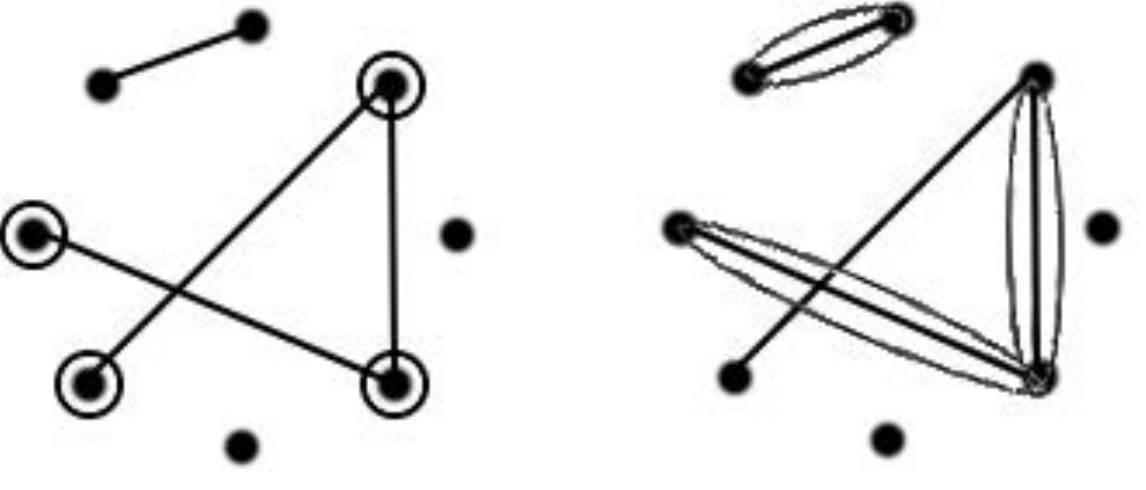}
\end{center}
\caption[]{Node-based network sampling and link-based network sampling}
\end{figure}
The link-based method is the most suitable when investigating certain characteristics of links in social network services. Nevertheless, it is shown that the link-based sampling method just preserves some of the topological structures of a social network (Lee et al., 2006). Besides, it shares the same weakness with node-based sampling, as random sampling on links in online social networks is often not an easy task.

\subsection{Traversal Network Sampling}
When conducting sampling on a social network, it is convenient to use the property that the nodes are connected with each other. Therefore, we can refer from one node to another repeatedly using the existing links between them, which can accelerate the speed of sampling on SNSs. Methods of this type are generally called chain-referral methods, almost all of which are biased towards nodes with higher degree. While the bias for chain-referral sampling with replacement is known asymptotically, the bias for chain-referral sampling without replacement, i.e. traversal sampling, has no direct formula, approximations of which will be provided in Section 3. The following provides an account for common traversal network sampling methods.

The Breadth-First-Search (BFS) sampling, i.e. the snowball sampling method, starts with a certain node in a social network and then samples all its network neighbors, and then all the neighbors of its network neighbors, etc. until the total number of nodes reach a certain amount. The BFS sampling is arguably the fastest method of all, and it is also feasible in current online social networks. However, evidence has shown that the BFS sampling tends to distort the topological features of a network, and it also suffers from other problems such as low convergence rates and very high bias towards high-degree nodes. The traversal sampling methods below share similar properties with BFS (Kurant et al., 2011b).

The Depth-First-Search (DFS) sampling starts with a certain node and goes along a random route until it reaches a terminal. Then it retreats to the nearest visited node joining another branch and goes along this branch randomly until it reaches a terminal again, and so force until a whole connected component of a social network is visited. This method aims at visiting a whole network or one connected component, and when it is carried out incompletely, it also introduces an unknown bias towards high-degree nodes.

Forest Fire sampling is a modification of the BFS sampling. In Forest Fire sampling, we start with a certain node, then sample each of its neighbors with probability $p$ and then sample each of the neighbors of its sampled neighbors with probability $p$, etc. until we have collected a certain number of samples. Forest Fire sampling reduces to BFS when $p=1$.

An alternative of snowball sampling, simply referred to as snowball sampling in some texts (not to be confused with the BFS method), is another modification of the BFS method. Instead of sampling all neighbors of a node, we now just sample $n$ neighbors of a node at each step, and then reject the neighbors that we have already sampled. The reader can refer to Illenberger et al. (2009) for a crude estimation of the sampling probabilities for this approach which can be applied for bias correction.
\begin{figure}[h]
\begin{center}
    \includegraphics[width=3in,height=1in]{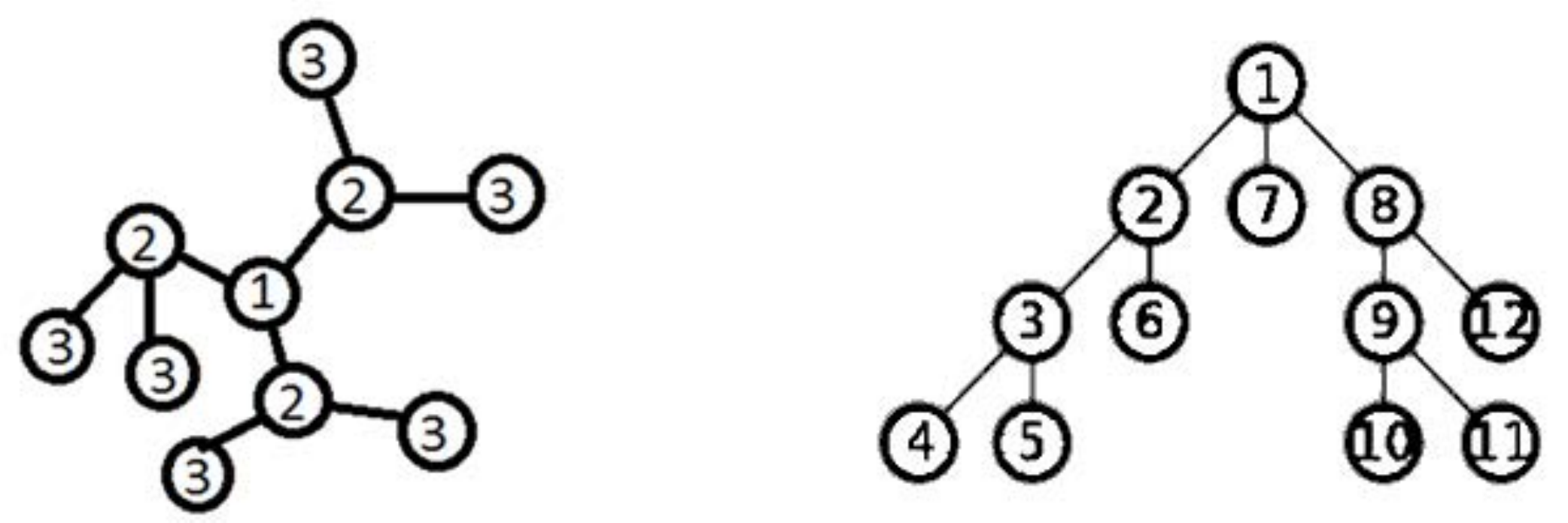}
\end{center}
\caption[]{A comparison between BFS sampling and DFS sampling}
\end{figure}

\subsection{Random Walk Network Sampling}
Random walk (RW) sampling means one-node-at-a-time chain-referral sampling on an SNS with replacements. It possesses many desirable properties such as fast convergence and known asymptotic sampling probabilities, although it might not be useful in terms of non-local graph properties like the graph diameter and the average shortest path length (Gjoka et al., 2010). Moreover, from the further analysis in Section 4, in terms of sampling error, random walk may not work as well as traversal methods as the former samples nodes repetitively. The sampling results should be re-weighted with the Hansen-Hurwitz estimator in Section 4.1 or the like. The variations of random walk sampling are listed as follows:

Simple random walk visits at each step one of the neighbors of the previous node with the same probability. If the network is connected and aperiodic, the sampling probability of a particular node $v$ converges to the stationary distribution: $\pi(v)=\deg(v)/2\vol(V)$, $V$ being the whole network.

Metropolis-Hastings random walk visits at each step the neighbor $v$ of the last node $u$ with the same probability and sample the new node with probability $\min(1,\deg(u)/\deg(v))$. Otherwise, we sample the node $u$ again. Then the sampling probability of each node is the same: $1/|V|$.

Weighted random walk places weight $w(u,v)$ on each link $(u,v)$ of a network so that the desired nodes incur more weight. Then we visit at each step the neighbor $v$ of the last node $u$ with probability $w(u,v)/(\sum_{v\in N(u)}w(u,v))$, $N(u)$ being the neighborhood of $u$. The sampling probability of a node $v$ is then $\pi(v)=w(v)/(\sum_{u\in V}w(u))$, $w(v)=\sum_{v'\in N(v)}w(v,v')$. For sampling on a network with disproportionate groups, the reader can refer to Kurant et al. (2011a) for an intricate algorithm for calculating the weights in order to acquire proportionate sample volumes for each group.

\subsection{Respondent-driven Network Sampling}
Respondent-driven sampling (RDS) was first invented as a chain-referral sampling method to detect hidden populations (Heckathorn 1997) and nowadays it generally refers to chain-referral sampling methods which sample $n$ nodes from one node at a time. Both with-replacement and traversal sampling are included and RDS may also comprise multiple starts, which lead to multiple chains of samples. When $n = 1$, RDS with replacements reduces to simple random walk and when RDS is performed without replacements, it can be viewed as the alternative of snowball sampling above. Even without replacements, the sampling probabilities of RDS can be considered as proportional to the node degree. Still, this approximation is too crude and two ways to improve it are explained below.

\section{Estimation of the Sampling Probabilities}
As is mentioned above, for with-replacement chain-referral methods, we know that the sampling probability $\pi(v)\doteq\deg(v)/2\vol(V)$. For traversal sampling, Kurant et al. (2011b) provide a direct estimation of the sampling probabilities. They view traversal sampling as a process randomly connecting the ``stubs'' inside a network, a degree-$k$ node with $k$ stubs. They assign each stub with an index $\iid U(0,1)$, and assume that there is a time process from $t=0$ to $t=1$. A node is connected when $t$ reaches the smallest index of its stubs. Then given fixed time $t$, $\pi(v)\propto 1-(1-t)^{\deg(v)}$, and the sampling proportion $f \approx Ef = 1-\sum_k p(k)(1-t)^k$, which we denote as $g(t)$, $p(k)$ being the proportion of degree-$k$ nodes. Therefore $\pi(v) \propto 1-(1-g^{-1}(f))^{\deg(v)}$.

The node degree distribution $p(k)$ as above needs estimation, and Kurant et al. (2011b) have provided a number of approaches, the most direct of which is to run a random walk on the network and estimate $p(k)$. This approach will be adopted below, and the author also proposes to further estimate $g(t)$ as $1-(1-t)^{\sum p(k)k} = 1-(1-t)^{\bar{d}}$, $\bar{d}$ being the average node degree. Taking traversal sampling as with-replacement sampling, $\bar{d}$ should be the harmonic average node degree applying the Hansen-Hurwitz estimator below. It was discovered during simulation that the arithmetic average node degree resulted in little difference, which was eventually applied for simplicity. The whole proposal might be bold, yet later it will be shown that the proposal is reasonable most of the time, and it mitigates the computational burden as well.

Gile (2011a) provides a successive sampling (SS) estimation for the problem based on bootstrapping. It is first assumed that $\pi_0(v)\propto\deg(v)$, as in sampling with replacements. Given an estimation of sampling probability $\pi_{i-1}(v)$, the number of degree-$k$ nodes in the population $V_k^i \approx |V|\frac{v_k}{\pi_{i-1}(v)}/ \sum\frac{v_k}{\pi_{i-1}(v)}$, $v_k$ being the number of degree-$k$ nodes in the sample. We generate a random network with volume $|V|$ and number of degree-$k$ nodes $V_k^i$, and then draw $M$ successive samples with original sample size from it. The estimation of sampling probability can be updated as $\pi_i(v)=(U_k+1)/(MV_k^i+1)$, $U_k$ being the total number of degree-$k$ nodes in the $M$ samples. This procedure is repeated till a satisfactory $\pi_r(v)$ is reached. It should be noticed that the prior knowledge of the population size is required. It has shown in the original paper that this estimation outperforms with-replacement RDS. In the next section the author will reconfirm the result while providing a more detailed comparison of network sampling techniques.

\section{Statistical Inference for Sampling SNSs}
In this section, we will discuss statistical inference for sampling social networking services. Regarding what should be estimated, we simulate networks with population size $1000$ and randomly assign each node with category A or B. The proportion of category A could be controlled to be, say, $0.3$ in the following examples. Then variations of RDS will be applied to estimate the proportion of category A in terms of point estimation discussed in Section 4.1 and confidence interval estimation discussed in Section 4.2. Due to the scarcity of real data, all simulation results are based on ERGMs (Snijders, 2002) and averaged over $100$ random networks with population size $1000$ and $100$ samples on each network without further specification.

\subsection{Point Estimation of Node Properties}
Because RDS methods are all biased towards higher-degree nodes, we cannot use the category proportions of the sample to replace the category proportions of the total population. The Hansen-Hurwitz (or generalized Horvitz-Thompson, Volz-Heckathorn) estimator (Hansen and Hurwitz, 1943; Gile and Handcock, 2010) estimates population total as $\hat{x}_{total}=\frac{1}{n}\sum_{v\in V}\frac{x(v)}{\pi(v)}$, and population mean as $\bar{x}=\sum_{v\in V} \frac{x(v)}{\pi(v)}/\sum_{v\in V}\frac{1}{\pi(v)}$; $x(v)$ is the node characteristic and $\pi(v)$ is the sampling probability. This estimator is asymptotically unbiased and is applied for RDS in most situations. For an alternative, the reader can apply the classic Horvitz-Thompson estimator $\bar{x}=\sum_{v\in V} \frac{x(v)}{\pi(v)}/(|V|n)$, $n$ being the sample size. We need only make $x(v)$ an indicator function to estimate the category proportions of the total population.

\subsubsection{Sampling with Replacements}
Now we investigate the accuracy of random walk and RDS with replacements and estimate the proportion of category A of social networks. We use the relative mean error as an indicator which is calculated from a sample of volume $n$ on a quadratic mean basis: $\hat{e}_r=\hat{e}/p$, $\hat{e} =\sqrt{\sum e_i^2/n}$, $e_i$ being each absolute error and $p$ being the category proportion. The relative mean errors are plotted below against the sampling proportions to show how many samples we should collect in order to obtain a reliable estimation. There are comparisons between random walk and RDS when $n=3$.

To identify the influence of network structure on sampling results, we control two parameters, namely the homophily ratio and the activity ratio. The former indicates the tendency of nodes to be connected within the categories, and is calculated by dividing the expected number of ties between category A and category B in networks with the same average node degree by the actual number of ties between category A and category B. The activity ratio is defined to be the ratio between the average node degree of category A and the average node degree of category B. The figures below in the upper row are produced from networks with homophily ratio 1, which indicates normal networks; those in the lower row are produced from networks with homophily ratio 2, indicating more ties within categories. From the left to the right, the activity ratio is 0.5, 1 and 2. The figures are presented below:

\begin{figure}[h]
\begin{center}
    \includegraphics[width=6in,height=3.6in]{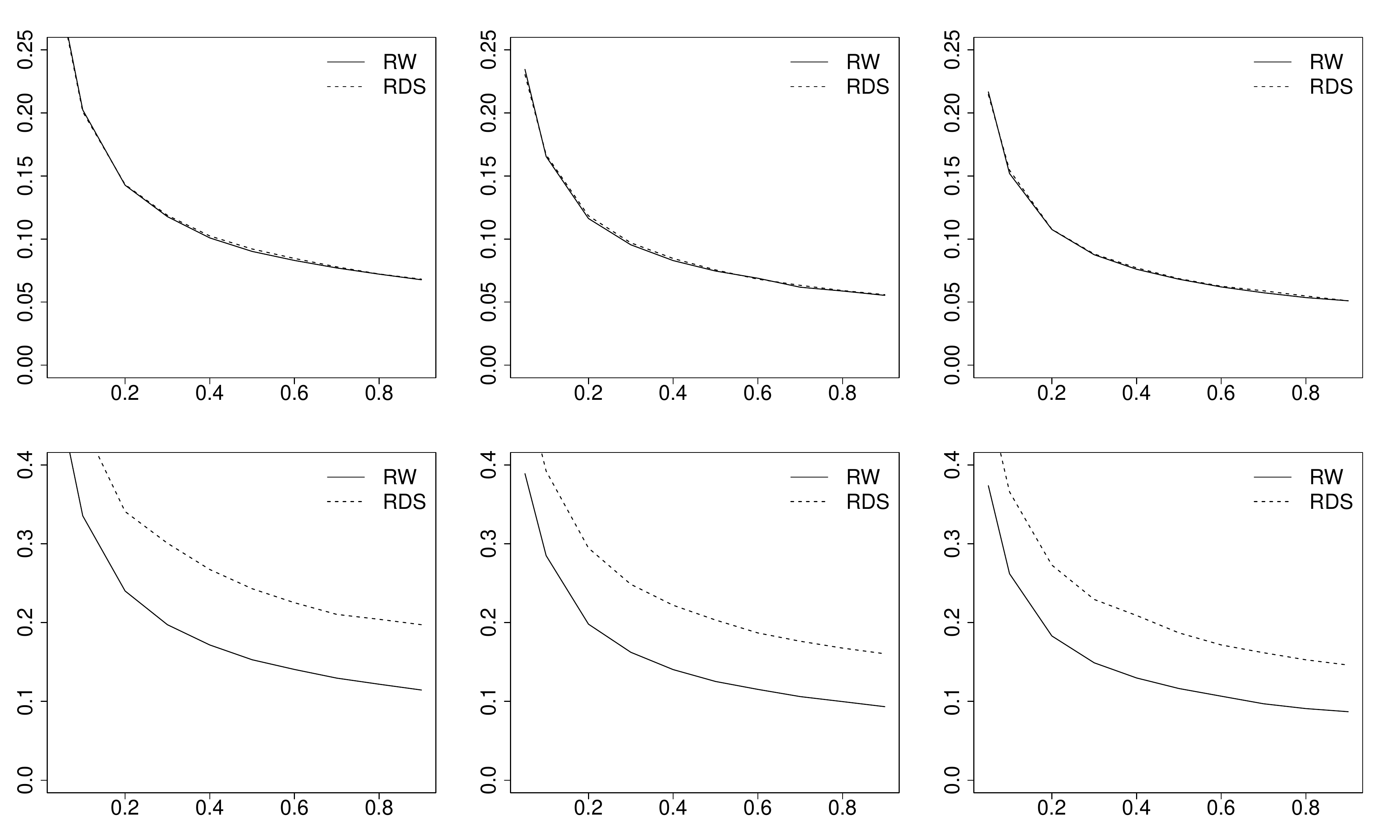}
\end{center}
\caption[]{The relative mean errors of with-replacement RW and RDS (n=3) against sampling proportion on different social networks. From left to right: activity ratio = 0.5, 1, 2; from top to bottom: homophily ratio = 1, 2}
\end{figure}

All social networks of different activity ratios and homophily ratios indicate that the mean error generally decreases with more sampled nodes, though fast at first and quite slow later. Also, the patterns of RW and RDS curves are highly similar so they almost coincide and no one seems to clearly outperform another when there is no homophily. However, when the homophily ratio is large, RW tends to outperform RDS to a large extent, and all errors tend to be much larger. The sampling error decreases with an increase in activity ratio, which is consistent with the fact that a network with dense ties is easier to sample than one with fewer ties. Meanwhile, there is a tendency that the mean error does not decrease rapidly with an increase in sample volume. In contrast, they seem to die down to a certain positive number (in these examples, around 0.05 and 0.10) with the sample volume increasing, though they will eventually go to zero. Moreover, sampling proportions larger than one are particularly undesirable.

It was also found in simulation that when the homophily ratio is high, there are some outliers which significantly increase the mean error. This is likely to be caused by the fact that the sampler is trapped in a small group of people with dense ties. This phenomenon should be avoided in both theory and practice, and the most natural solution seems to be avoiding with-replacement sampling if possible.

It is reasonable to hypothesize that the mean error does not quickly converge to zero as sample size increases because the sampler picks up some nodes with extreme properties, which increases the sample variation of the node characteristic estimation to a large extent. The following simulation results, with sampling errors decomposed into sample bias and sample standard deviation, largely prove this hypothesis, which are shown as follows:

\begin{figure}[h]
\begin{center}
    \includegraphics[width=6in,height=1.8in]{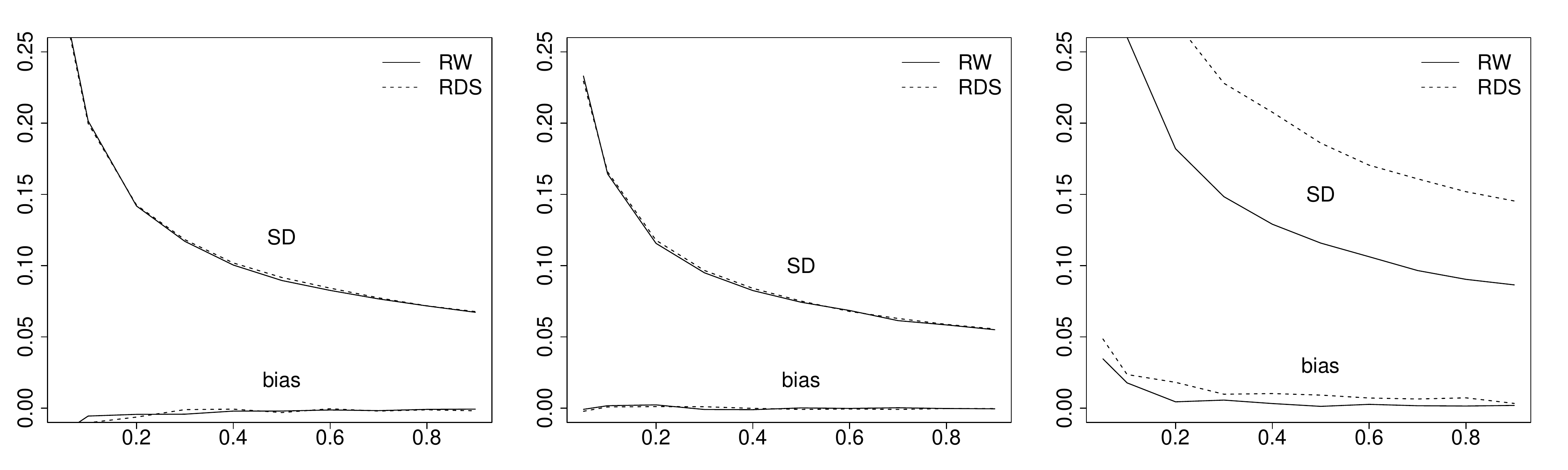}
\end{center}
\caption[]{The relative sample biases and sample standard deviations of with-replacement RW and RDS (n=3) against sampling proportion on different social networks. From left to right: homo. ratio = 1, act. ratio = 0.5; homo. ratio = 1, act. ratio = 1; homo. ratio = 2, act. ratio = 2.}
\end{figure}

We can observe from the figures above that the sampling bias is almost zero, and the main contributor to the error is the standard deviation. The standard deviation decreases rather slowly as the sampling proportion increases, which explains the slow convergence of with-replacement sampling on a social network. Meanwhile, the convergence of the traversal sampling is guaranteed as the sampling proportion tends to one. We will compare the results above with the results from traversal sampling below.

\subsubsection{Traversal Sampling}
The following figures exhibit the relative mean errors of category proportion estimated by with-replacement sampling and traversal sampling. The sampling probabilities of traversal sampling are estimated by both methods in Section 3.

\begin{figure}[h]
\begin{center}
    \includegraphics[width=6in,height=1.8in]{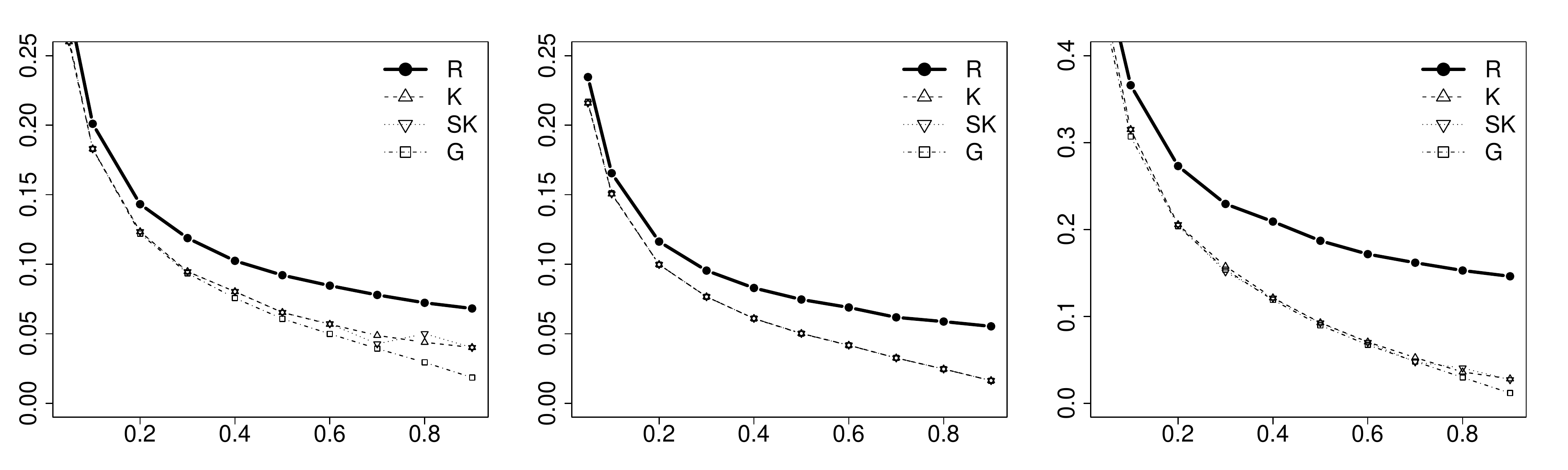}
\end{center}
\caption[]{The relative mean errors against sampling proportion on different social networks. Left: homo. ratio = 1, act. ratio = 0.5, RDS(n=3); middle: homo. ratio = 1, act. ratio = 1, RW; right: homo. ratio = 2, act. ratio = 2, RDS(n=3). R: with-replacement sampling; K: Kurant et al.'s direct estimation; SK: Kurant et al.'s direct estimation (simplified); G: Gile's successive sampling (SS) estimation. Other cases in Figure 3 have identical patterns with any of the above cases and are not listed.}
\end{figure}

It can be referred from the results that all estimates from traversal sampling consistently have smaller average errors than with-replacement estimates, which shows more applicability of traversal sampling. In the traversal estimates, Gile's SS estimates consistently perform better, but the difference is only significant when the sampled proportion is large (above 0.4) and the activity ratio is far from one, which is not very common in real practice. Considering the fact that Kurant et al.'s direct estimates are much easier to calculate than Gile's SS estimates, it is acceptable to apply the direct estimates under ``normal'' circumstances, where the sampling proportion and the requirement for accuracy are not too high. Also, it is shown that the simplified direct estimates almost have the same results as original direct estimates.

Further investigation can be made into the similarity between the Gile's SS estimates and the direct estimates. The correlation factors between the two estimates under all different circumstances in Figure 3 have been calculated and the numbers are all above 0.95 for sampling proportion $\le$ 0.7. When the sampling proportion > 0.7 and activity ratio is far from 1, the correlation factors drop below 0.95 but still are greater than 0.5. Therefore, the two estimates are also interchangeable under ``normal'' circumstances.

A temporary conclusion is that traversal methods work better with approximate bias corrections, while with-replacement sampling generally yields larger errors. It is an interesting question why the mean errors of Gile's SS estimates and Kurant et al.'s direct estimates often coincide, although they are computed in totally different ways. Once we identify the reason we may be able to maintain the accuracy of Gile's SS estimates with simpler computations.

\subsubsection{Relationship Between Error and Population Size}
Because online social networks are often of very large size, it is beneficial to discuss the change of the error when the population size increases. We can see from below that the errors are quite stable when the population size increases and the sample size keeps the same. The following table indicates the relative standard errors for the category proportion estimated by traversal RDS ($n=3$) with simplified Kurant et al.'s direct method and Gile's SS method when the population size equals 500, 700, 1000, 1300, 1700, 2000 and the sample size remains 100:

\begin{table}[h]
  \centering
$$
\begin{array}{|c|c|c|c|c|c|c|}
  \hline
  {\rm Population} & 500 & 700 & 1000 & 1300 & 1700 & 2000 \\ \hline
  {\rm Gile's\ SS} & 14.1\% & 14.6\% & 15.0\% & 15.1\% & 15.2\% & 15.3\% \\ \hline
  {\rm Kurant} & 14.1\% & 14.6\% & 15.0\% & 15.1\% & 15.2\% & 15.35\% \\ \hline
\end{array}
$$
  \caption{Sampling errors under different population sizes; homo. ratio = act. ratio = 1}
\end{table}

From these results, we can observe that the sampling errors increase quite slowly with population size increasing, and the increasing trend tend to diminish as population increases. Therefore, the results obtained are quite stable, which provide some evidence that the earlier results will still hold for networks with larger sizes as well. Still, considering the very complicated structure of networks, there is still much to be investigated in this area.

\subsection{Confidence Interval Estimation of Node Properties}
Constructing confidence intervals for node properties from network samples should be another issue for consideration. While ``naive'' confidence intervals\footnote{The ``naive'' confidence interval is the confidence interval with formula $[\bar{x}-t_{\al/2,n-1}\hat{s}/\sqrt{n}, \bar{x}+t_{\al/2,n-1}\hat{s}/\sqrt{n}]$.} generally do not perform satisfactorily according to Salganik (2006), there are currently two types of bootstrap confidence intervals for node properties which are acceptable for respondent-driven sampling on networks.

\subsubsection{Available Confidence Intervals}
The first type is Salganik's bootstrap confidence interval (2006). We first gather a sample from a network with RDS and then resample on the sample with replacements for $N$ times. The resampling procedure is that if the last resampled node is of category $A$, we resample the next node from all sampled nodes referred by category $A$, and so forth. Then we calculate the estimations of the node attributes $\hat{x}_1, \ldots, \hat{x}_n$ from e.g., the Hansen-Hurwitz estimator for each of the resampled samples and construct the confidence interval using $[\bar{x}-z_{\al/2}\hat{s}_e, \bar{x}+z_{\al/2}\hat{s}_e]$; $\bar{x}=\sum\hat{x}_i/N$ and $\hat{s}_e=\sqrt{\sum(\hat{x}_i-\bar{x})^2/(N-1)}$.

The second type is Gile's successive sampling confidence interval (2011b) which takes the homophily ratio and the activity ratio into account. In this method, we first estimate the node degree distribution of the network and simulate a network with same population and estimated node degree distribution. Then we denote the current number of nodes with category $i$ and degree $k$ as $\bar{N}_{i,k}$ and the number of links from category $i$ to category $j$ as $H_0(i,j)$. The number of links within and between categories are estimated as $H_0(A,A)=\bar{d}_A\sum_k\hat{N}_{A,k}\hat{r}_A$, $H_0(B,B)=\bar{d}_B\sum_k\hat{N}_{B,k}(1-\hat{r}_B)$ and $H_0(A,B)=H_0(B,A)=[\bar{d}_A\sum_k \hat{N}_{A,k}(1-\hat{r}_A)+\bar{d}_B\sum_k\hat{N}_{B,k}\hat{r}_B]/2$; $\bar{d}_i=\sum_k\hat{N}_{i,k}k/\sum_k\hat{N}_{i,k}$ and $\hat{r}_i$ is the sample proportion of links from $i$ to $A$ in all links from $i$. We sample from the simulated network just as sampling from the original network with the following difference: from a category-$i$ node, we first choose category $j$ with weight $H(i,j)=H_0(i,j)\cdot$proportion of unsampled category-$j$ nodes and then sample from {\it all} category-$j$ nodes with weights proportional to the node degrees. The sampling procedure is repeated $N$ times and the confidence interval can be constructed similarly to the last paragraph.

\subsubsection{The Homogeneity Between Different Networks}
In the above, Gile (2011b) has applied the method of simulating one network for standard deviation calculation instead of taking an average on different networks. To know whether the main variance of estimates comes from within the networks or between the networks, we perform an ANOVA analysis on category proportion estimation of $20$ networks with size $1000$ and $25$ RDS ($n=3$) samples with size $100$ on each network, and estimate the category proportions from the gathered samples with the same assumptions as before. The result shows that the right-tail probability of the $F$ statistic is $0.3725$. Therefore, it is reasonable to hypothesize that there is not a significant structural difference for sampling between different networks, and Gile's approach is reasonable and could be continued.

\subsubsection{A Faster Confidence Interval}
While the above confidence intervals aim at imitating the process of sampling a network with certain attributes, the proposed confidence interval intends to simulate a network with certain attributes and then sample on it. To achieve this goal, we simulate a network with given average node degree, homophily ratio and activity ratio, which could be estimated as follows.

For the average node degree $\bar{d}$, we can apply the Hansen-Hurwitz estimator and direct estimates or Gile's SS estimates as illustrated before. To estimate the homophily ratio, we assume that each link is sampled with the same probability, which is asymptotically true for with-replacement sampling. The estimated proportion of links between categories is then $C_{b}/C$, $C$ being the number of sampled links and $C_{b}$ being the number of sampled links between categories. The average proportion of links between categories for a general network is $\frac{n_An_B}{|V|(|V|-1)/2}$, and therefore we can estimate the homophily ratio as $\hat{h} = \frac{Cn_An_B}{C_b|V|(|V|-1)/2}$. The estimated activity ratio is $\hat{a}=\bar{d}_A/\bar{d}_B$, $\bar{d}_A$ and $\bar{d}_B$ being the estimated average node degrees of of category A and B using direct estimates or Gile's SS estimates.

From these results, we can now construct a faster confidence interval for respondent-driven sampling with the following steps:

(1) Simulate a network with the same sample size and population, estimated mean degree $\bar{d}$, homophily ratio $\hat{h}$ and activity ratio $\hat{a}$;

(2) Sample from the network $N$ times with the same sampling method and create $N$ samples of the estimated category proportion or some other node attribute;

(3) Compute the sample variance using the values of the estimates and hence construct the desired confidence interval as before.

To investigate on the performance of different confidence intervals, their coverage probabilities ({\it CP}s) are potentially useful. Moreover, to provide more information, the estimated standard deviations ({\it SD}s), which are related to the interval lengths, are also presented. The following is the performance of different kinds of $95\%$ bootstrap confidence intervals using traversal RDS ($n=3$), based on an average of the results from simulated networks. For case A, homophily ratio = 1, activity ratio = 1, sampling proportion = 0.1; for case B, homophily ratio = 1, activity ratio = 2, sampling proportion = 0.2; for case C: homophily ratio = 2, activity ratio = 2, sampling proportion = 0.3. Kurant et al.'s directed estimates (simplified) are applied for the new confidence intervals.
\begin{table}[h]
  \centering
$$
\begin{array}{|c|c|c|c|c|c|c|}
\hline
{\rm Case} & CP {(\rm Salganik's)} & SD {(\rm Salganik's)} & CP {(\rm Gile's\ SS)} & SD {(\rm Gile's\ SS)} & CP {(\rm new)} & SD {(\rm new)} \\ \hline
{\rm A} & 94.8\% & 0.046 & 94.0\% & 0.045 & 95.0\% & 0.046 \\ \hline
{\rm B} & 96.1\% & 0.032 & 94.2\% & 0.030 & 94.4\% & 0.028 \\ \hline
{\rm C} & 91.3\% & 0.041 & 93.2\% & 0.043 & 89.2\% & 0.042 \\ \hline
\end{array}
$$
  \caption{Performance of different kinds of $95\%$ bootstrap confidence intervals}
\end{table}

It can be observed that when the characteristics of networks such as activity ratio and homophily are not significant, the new confidence intervals tend to outperform the original ones. However, when such properties are significant, especially homophily, the new confidence intervals tend to be more inaccurate. The fluctuations in the coverage probabilities indicate the complexity of this problem. Still, the three sets of confidence intervals do not differ by a great amount in terms of coverage probabilities. Therefore, the new confidence intervals can be applied when the sampled network does not have specific characteristics, or the requirement for accuracy is not high and faster computational speed is desired.

Moreover, it would also be beneficial to investigate the change in the performance of confidence intervals with adjustments in calculation, where a number of questions could be raised. For instance, we have assumed that each link with sampled with equal probability, which is only true for with-replacement sampling. This could be one point for improvement. Meanwhile, it is also unknown whether it is better to sample across the whole network as in Salganik's and Gile's confidence intervals, or apply traversal sampling as in the new ones. It is probable that the latter results in larger sample standard deviation and more conservative estimates, although they are not reflected in the table above. Apart from that, to which extent we should replicate the network exactly as the original one is also a question, as more complicated replication does not necessarily result in better results, as is illustrated in case B. The new confidence interval is admittedly subject to much improvement, but because of computational burden, the author is regretful for being unable to present further results, which can be left to the reader to explore.

The confidence intervals above are potentially useful for academic researchers in online surveying, where RDS-type sampling methods are very likely to be suitable. It would not be difficult to apply these results on popular social networking services, e.g. Facebook and Twitter, to investigate the proportion of a certain group in a population. The author is also regretful that such an example is unavailable here, which might be constructed by readers in need.

\section{Discussion and Related Works}
There have been numerous articles on social network analysis, and many of them have been devoted to social network sampling. For traversal sampling, Kurant et al. (2011b) provide a comprehensive list of sampling techniques and an approach to correct the bias as is mentioned before. For random walk methods, Gjoka et al. (2010) provide a comparison between simple random walk and Metropolis-Hastings random walk, and Kurant et al. (2011a) provide a stratified sampling technique for sampling disproportionate categories. The reader may also refer to Stutzbach et al. (2006) for a modified MHRW technique for sampling dynamic networks. In terms of respondent-driven sampling, the reader can refer to Heckathorn (1997; 2002) and Salganik and Heckathorn (2004) for some early discussion on RDS and the RDS-I type estimator. The RDS-I type estimator has become outdated as its performance is relatively inferior to the RDS-II type estimator (Volz and Heckathorn, 2008), which is basically the Hansen-Hurwitz estimator. Gile and Handcock (2010) provide an examination of the RDS techniques as are mentioned above. Wejnert and Heckathorn (2008) provide a example of RDS for online social networks, and Rasti et al. (2009) show an example of RDS for sampling dynamic networks. While a vast number of methods have been proposed for social network sampling and related estimation, new methods with higher efficiency may still be invented in the future.

To evaluate the performance of chain-referral sampling methods, there has been much discussion on the mixing time of the Markov chain. Sinclair (1992) provides theoretical evidence that the mixing time is bounded by the second largest eigenvalue of the transition matrix of the Markov chain. For empirical evidence, the reader can refer to Mohaisen et al. (2010) which shows that the mixing time of many chains is higher than expected. Mohaisen et al. (2012) provide a comprehensive empirical analysis of the mixing speed of BFS, RW and MHRW with examples of sampling online social networks. To accelerate the mixing speed of the Markov chain, there have been a number of proposed techniques, such as uniform restarts (Avrachenkov et al., 2010), multiple starts (Ribeiro and Towsley, 2010) and a fastest mixing algorithm based on the network structure (Boyd et al., 2004). On the other hand, there has been relatively limited literature on the performance of estimates from chain-referral sampling methods, some of which can be found in Gile and Handcock (2010) and Gile (2011a), and improving the performance of the estimates constitutes another question. The article above has provided some new evidence on the performance of sampling social networking services and has also proposed some new methods to improve the related estimates. Meanwhile, much still remains to be done.



\end{document}